\documentclass[pre,aps,amssymb,amsmath,superscriptaddress,twocolumn]{revtex4}
\usepackage{amsmath}
\usepackage{amssymb}
\usepackage{epsfig}
\usepackage{amscd}
\usepackage{graphicx,psfrag,xspace}
\usepackage{float}
\usepackage[normalem]{ulem}

\begin{document}

\title{Residence time and collision statistics for exponential flights: the rod problem revisited}
\author{A. Zoia}
\email{andrea.zoia@cea.fr}
\affiliation{CEA/Saclay, DEN/DANS/DM2S/SERMA/LTSD, 91191 Gif-sur-Yvette, France}
%\affiliation{Commissariat \`a l'Energie Atomique et aux Energies Alternatives, Direction de l'Energie Nucl\'eaire, D\'epartement de Mod\'elisation des Syst\`emes et Structures, Service d'Etudes des R\'eacteurs et de Math\'ematiques Appliqu\'ees, CEA/Saclay, 91191 Gif-sur-Yvette, France}
\author{E. Dumonteil}
\affiliation{CEA/Saclay, DEN/DANS/DM2S/SERMA/LTSD, 91191 Gif-sur-Yvette, France}
\author{A. Mazzolo}
\affiliation{CEA/Saclay, DEN/DANS/DM2S/SERMA/LTSD, 91191 Gif-sur-Yvette, France}

\begin{abstract}
Many random transport phenomena, such as radiation propagation, chemical/biological species migration, or electron motion, can be described in terms of particles performing {\em exponential flights}. For such processes, we sketch a general approach (based on the Feynman-Kac formalism) that is amenable to explicit expressions for the moments of the number of collisions and the residence time that the walker spends in a given volume as a function of the particle equilibrium distribution. We then illustrate the proposed method in the case of the so-called {\em rod problem} (a $1d$ system), and discuss the relevance of the obtained results in the context of Monte Carlo estimators.
\end{abstract}
\maketitle

\section{Introduction}

The so-called Pearson random walk describes the evolution of particles starting from a point-source and performing straight-line displacements until collision events, where either the direction of propagation changes at random with probability $p$ (scattering), or the trajectory is terminated (absorption)~\cite{hughes, weiss}. When the traversed medium is homogeneous, so that the scattering centers are uniform, the inter-collision distances (flights) are exponentially distributed. Exponential flights are key to understanding the dynamics of many transport processes, encompassing areas as diverse as radiation transfer, electron motion in semiconductors, gas dynamics, and search strategies~\cite{cercignani, wigner, jacoboni_book, lecaer, blanco_fournier, benichou_epl}. In most such applications, one is typically interested in assessing the particle density $\Psi_V$ (or some functional defined on $\Psi_V$) averaged over a $d-$dimensional volume $V$ in the phase space. In Reactor Physics, for instance, $\Psi_V$ might represent the number of particles escaping from radiation shielding~\cite{wigner}.

\begin{figure}[b]
\centerline{\epsfxsize=9.0cm\epsfbox{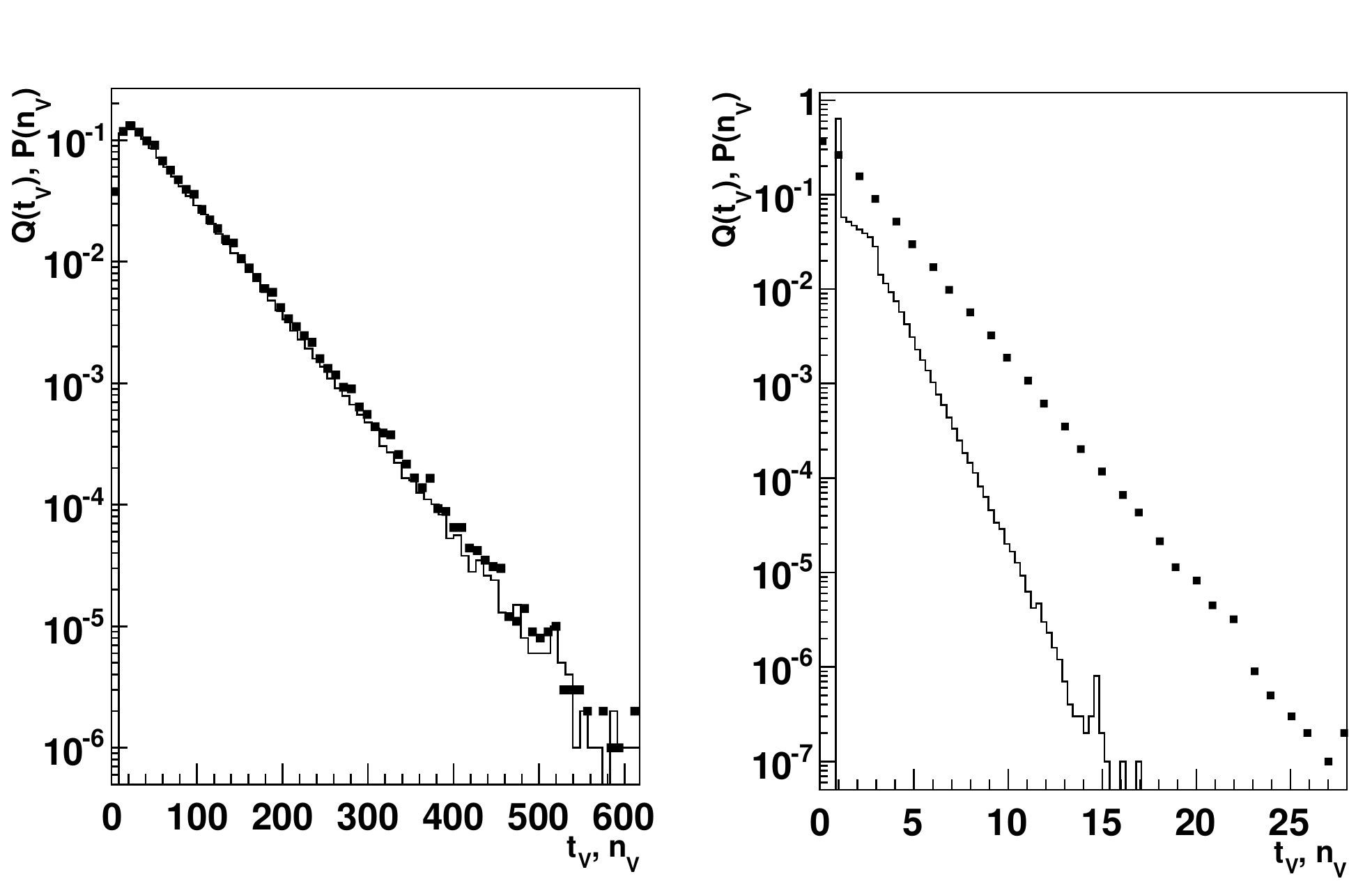}}
\caption{The normalized distributions of collision number ${\cal P}(n_{V})$ (dots) and residence times ${\cal Q}(t_{V})$ (solid line) in a volume $V$ for $1d$ exponential flights. Left. $R=10$ and $\lambda_t=1$. Right. $R=1$ and $\lambda_t=1$.}
   \label{fig1}
\end{figure}

The particle density of exponential flights, in turn, is intimately connected to the statistical properties of the collisions $n_{V}$ falling in the region $V$, and the residence time $t_V$ spent within $V$~\cite{benichou_epl, mazzolo, zoia_dumonteil_mazzolo, zdm_prl}. Even under simplifying hypotheses, namely, that scattering and absorption probabilities do not depend on particle energy, so that we can safely define an average speed $v$ (one-speed approximation), and that scattering is isotropic, the interplay between $\Psi_V$, $n_{V}$ and $t_V$ turns out to be a deceivingly simple problem, and has attracted a renovated interest in recent years~\cite{paasschens, orsingher, kolesnik, mazzolo, blanco, zoia_dumonteil_mazzolo, zdm_prl}. For instance, when the typical size $R$ of the volume $V$ is much larger than the mean free path $\lambda_t$ between collisions, namely, $R \gg \lambda_t$ (the so-called diffusion limit), the normalized distribution of collision number ${\cal P}(n_{V})$ and the normalized probability density of residence times ${\cal Q}(t_{V})$ converge to each other~\cite{zdm_prl}, as illustrated in Fig.~\ref{fig1} (left). This is in general not true when $R$ is comparable to $\lambda_t$, i.e., when finite speed effects and boundaries come into play, and particles spend only a limited number of collisions in $V$ before wandering away, as shown in Fig.~\ref{fig1} (right).

Aside from its theoretical interest for understanding the dynamics of exponential flights in bounded geometries, the study of ${\cal P}(n_{V})$ and ${\cal Q}(t_{V})$ is also motivated by their prominence in Monte Carlo methods. In view of the intrinsic stochastic nature of exponential flights, one is naturally led to resort to Monte Carlo simulation, which can guide the development of analytical solutions, and, in most realistic applications, provide the answers that are not accessible by analysis alone~\cite{lux, spanier}. In plain Monte Carlo methods, the volume-averaged particle density $\Psi_V$ for one-speed transport can be estimated by simulating particle trajectories in the phase space and either counting the collisions $n_{V}$ in $V$, or measuring the length $\ell_V$ of particle tracks within $V$. In the former case, we have $\Psi^{coll}_V=V^{-1} \lambda_t\sum_{i\in V} 1=n_{V}\lambda_t/V$, whereas in the latter $\Psi^{track}_V=V^{-1} \sum_{i\in V} \ell_i = \ell_V/V$~\cite{spanier}. As speed is assumedly constant, we can equivalently compute $\Psi^{track}_V$ by measuring the residence time $t_V=\ell_V/v$ that particles spend in $V$. It is well known that the two estimators described above are unbiased with respect to $\Psi_V$, which amounts to saying that $ \bar{\Psi}^{coll}_V \to \Psi_V$ and $\bar{\Psi}^{track}_V  \to \Psi_V$, where $\bar{\cdot}$ denotes averaging with respect to particle trajectories, and the limit is attained for an infinite number of realizations~\cite{spanier}. This in particular implies that the two estimators are related by $\bar{n}_V = \bar{t}_V / \tau_t$, where $\tau_t=\lambda_t/v$ is the average flight time. Such a property non-trivially holds for any kind of boundaries imposed on $V$ and stems from the memoryless (Markovian) nature of the underlying exponential flight process~\cite{lux, spanier}. Fig.~\ref{fig1} suggests that $t_V$ and $n_V$, while preserving the same average, will generally have different higher order moments, and in particular different variances. Hence, there might be an advantage in using either estimator for determining the desired particle density $\Psi_V$.

\begin{figure}[t]
\centerline{\epsfxsize=8.0cm\epsfbox{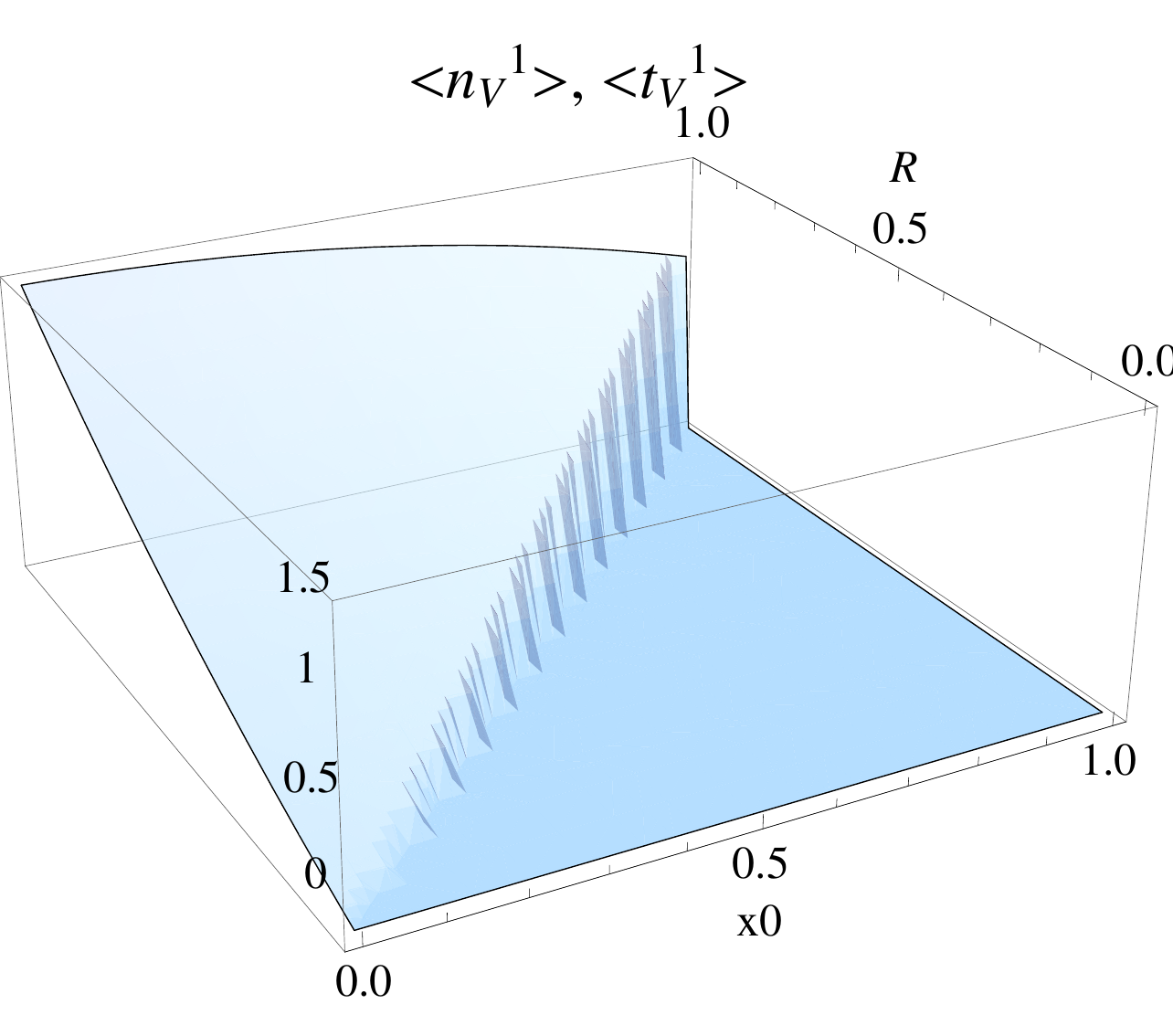}}
\caption{Mean first-passage time and collision number for leakage boundary conditions and $p=1$. The two surfaces coincide.}
   \label{fig2}
\end{figure}

In the following, we address the issue of characterizing the distribution of collisions $n_{V}$ and residence times $t_{V}$ in a volume $V$ for exponential flights. This paper is structured as follows. In Sec.~\ref{methodology}, we will first recall some preliminary background, and sketch a general approach for the moments of the distributions, based on the Feynman-Kac formalism. Then, in Sec.~\ref{rod_sec} we will exemplify the proposed methodology by explicitly evaluating those moments for a $1d$ domain, the so-called {\em rod problem}. Perspectives are finally discussed in Sec.~\ref{conclusions}.

\begin{figure}[t]
\centerline{\epsfxsize=8.0cm\epsfbox{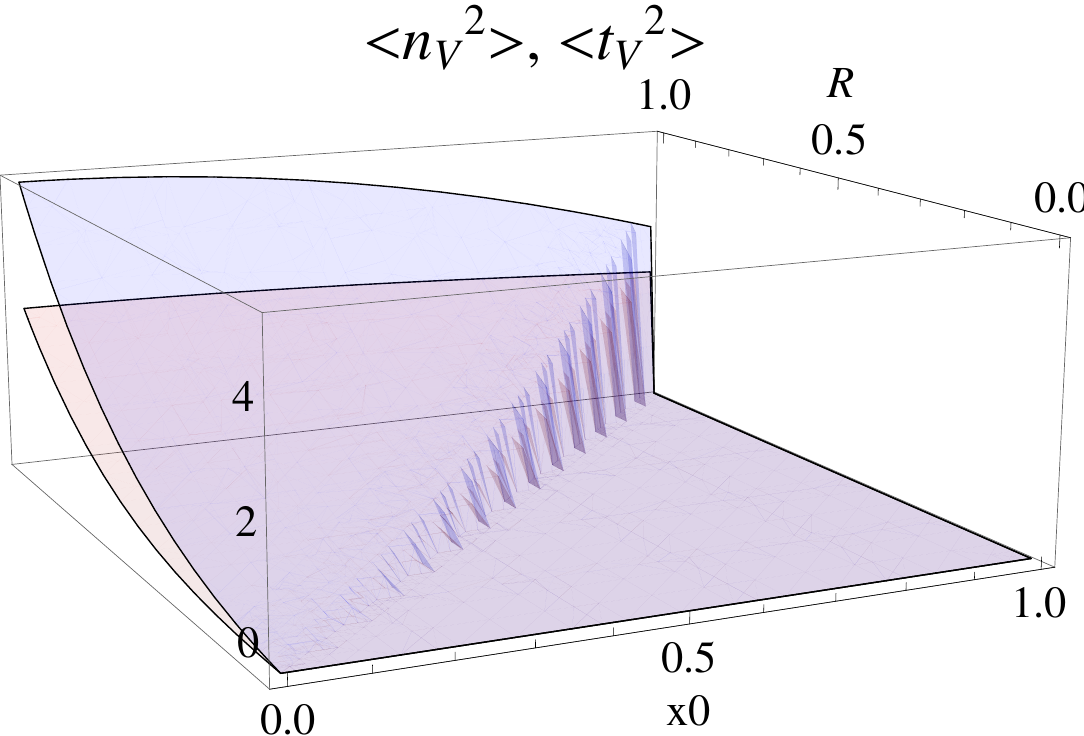}}
\caption{Second moment of first-passage time (dark surface) and collision number (light surface) for leakage boundary conditions and $p=1$.}
   \label{fig3}
\end{figure}

\section{Methodology}
\label{methodology}

The trajectory ${\mathbf z}_t=\left\lbrace {\mathbf r}_t,{\mathbf \omega}_t \right\rbrace$ of exponential flights in the phase space is defined by the stochastic evolution of position and direction, starting from the point-source ${\mathbf z}_0=\left\lbrace {\mathbf r}_0,{\mathbf \omega}_0\right\rbrace$ at time $t=0$. Due to the exponential nature of the displacement lengths, the stochastic process ${\mathbf z}_t$ is Markovian. In other words, knowledge of the pair position-direction at a given time enables to determine the system evolution~\footnote{This is generally not true for the process $\left\lbrace {\mathbf r}_t \right\rbrace$ alone, defining the direction-averaged position of the walker.}. For the sake of simplicity, we assume that scattering is isotropic. The propagator $\Psi({\mathbf r},{\mathbf \omega},t|{\mathbf r}_0,{\mathbf \omega}_0)$ defines the probability density for the walker being at a point $\left\lbrace {\mathbf r},{\mathbf \omega} \right\rbrace$ in the phase space, at a time $t$, having started from the initial condition. The propagator of exponential flights satisfies a probability balance, the forward Chapman-Kolmogorov equation
\begin{equation}
\frac{\partial }{\partial t}\Psi({\mathbf r},{\mathbf \omega},t|{\mathbf r}_0,{\mathbf \omega}_0) = {\cal L} \Psi({\mathbf r},{\mathbf \omega},t|{\mathbf r}_0,{\mathbf \omega}_0),
\label{chapman_kolmogorov}
\end{equation}
where ${\cal L}$ is the forward transport operator
\begin{equation}
{\cal L} = -{\mathbf v}\cdot \nabla_{\mathbf r} +\frac{1}{\tau_s} \int d{\mathbf \omega}-\frac{1}{\tau_t}.
\end{equation}
Here we have set ${\mathbf v}={\mathbf \omega}v$, $\tau_s=\lambda_s/v$, $\lambda_s$ being the scattering mean free path, and the integral over directions is normalized to $\Omega_d=2\pi^{d/2}/\Gamma(d/2)$, i.e., the surface of the unit sphere. We introduce then the collision density
\begin{equation}
\Psi({\mathbf r},{\mathbf \omega}|{\mathbf r}_0,{\mathbf \omega}_0)=\frac{1}{\tau_t}\int_{0}^{+\infty}\Psi({\mathbf r},{\mathbf \omega},t|{\mathbf r}_0,{\mathbf \omega}_0)dt,
\end{equation}
which intuitively represents the equilibrium distribution of the particle ensemble. Remark that the propagator $\Psi({\mathbf r},{\mathbf \omega},t|{\mathbf r}_0,{\mathbf \omega}_0)$ depends on the boundary conditions imposed on $\partial V$. The absence of boundary conditions corresponds to defining a fictitious (`transparent') volume $V$, where particles can indefinitely cross $\partial V$ back an forth. On the contrary, the use of leakage boundary conditions leads to the formulation of first-passage problems~\cite{redner, benichou, majumdar_fptd, condamin}, where the walker is lost upon crossing $\partial V$.

\begin{figure}[t]
\centerline{\epsfxsize=8.0cm\epsfbox{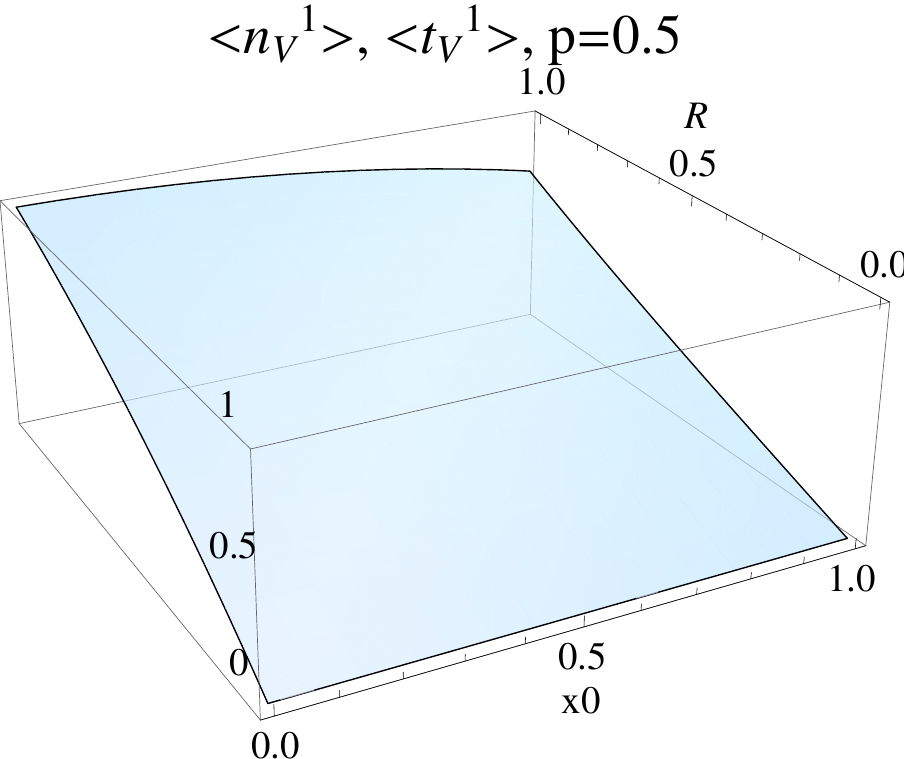}}
\caption{Mean residence time and collision number for transparent boundaries and $p=0.5$. The two surfaces coincide.}
   \label{fig4}
\end{figure}

The collision number $n_{V}$ and the sojourn time $t_V$ of the walker inside $V$ depend on the realizations of the trajectories ${\mathbf z}_t$, and as such are random quantities, whose behavior can be fully characterized in terms of their respective moments
\begin{eqnarray}
\langle n_V^m \rangle({\mathbf r}_0,{\mathbf \omega}_0) = \sum_{n_{V}=1}^{+\infty} n_{V}^m {\cal P}(n_{V}|{\mathbf r}_0,{\mathbf \omega}_0)\nonumber\\
\langle t_V^m \rangle({\mathbf r}_0,{\mathbf \omega}_0) = \int_{0}^{+\infty} t_{V}^m {\cal Q}(t_{V}|{\mathbf r}_0,{\mathbf \omega}_0) d t_{V}.
\label{moments_formulae}
\end{eqnarray}
Remark that ${\cal P}(n_{V}|{\mathbf r}_0,{\mathbf \omega}_0)$ and ${\cal Q}(t_{V}|{\mathbf r}_0,{\mathbf \omega}_0)$ depend on the initial conditions. Knowledge of all moments suffices to describe the associated distributions. In the following, we derive explicit expressions that allow evaluating the moments $\langle n_V^m \rangle({\mathbf r}_0,{\mathbf \omega}_0)$ and $\langle t_V^m \rangle({\mathbf r}_0,{\mathbf \omega}_0)$ in terms of the equilibrium distribution $\Psi({\mathbf r},{\mathbf \omega}|{\mathbf r}_0,{\mathbf \omega}_0)$.

\subsection{Residence times}
\label{time_sec}

In a series of seminal works based on Feynman path-integral formalism, Kac~\cite{kac_original, kac_berkeley, kac, kac_darling} has worked out a general method for deriving the residence time distribution when the underlying stochastic process is a Brownian motion $W_t$, and later showed that his results hold more generally for Markov processes~\footnote{Actually, the name {\em residence time} has been introduced in the Physics literature later on, by~\cite{agmon_original}. In the original papers, residence time was rather called {\em occupation time}, which has stuck in the Mathematics literature~\cite{kac_darling}.}. For a review (focused on Brownian motion), see, e.g.,~\cite{majumdar_review}. When a trajectory ${\mathbf z}_t$ is observed up to a time $t$, the associated residence time $t_V(t) \le t$ in $V$ is formally
\begin{equation}
t_V(t)=\int_{0}^{t} \chi[{\mathbf z}(t')] dt',
\label{res_time_definition}
\end{equation}
$\chi[{\mathbf z}]$ being the marker function of the domain $V$, which is equal to $1$ when ${\mathbf z} \in V$, and vanishes elsewhere. When $V$ has leakage boundary conditions, $t_V(t)$ for an infinite observation time corresponds to the first-passage time to the boundary $\partial V$. More generally, the definition in Eq.~\eqref{res_time_definition} allows for multiple exits and re-entry crossings of $\partial V$~\cite{benichou_epl}. The key ingredient of Kac approach is the stochastic integral $F(t,s|{\mathbf z}_0)=\langle e^{-s t_V(t)}\rangle$, where the expectation is taken with respect to the propagator $\Psi({\mathbf z},t|{\mathbf z}_0)$, i.e., the probability density of performing a trajectory from ${\mathbf z}_0$ at $t'=0$ to ${\mathbf z}$ at time $t'=t$, namely,
\begin{equation}
\langle e^{-s t_V(t)}\rangle=\int \Psi({\mathbf z},t|{\mathbf z}_0)e^{-s t_V(t)}d{\mathbf z}.
\label{stochastic_integral}
\end{equation}
The existence and well-posedness of Eq.~\eqref{stochastic_integral} is discussed in, e.g.,~\cite{kac_original, kac}. By slightly adapting the treatment in~\cite{kac}, it can be shown that $F(t,s|{\mathbf z}_0)$ satisfies the equation
\begin{equation}
\frac{\partial}{\partial t}F(t,s|{\mathbf z}_0)={\cal L}^* F(t,s|{\mathbf z}_0)- s\chi[{\mathbf z}_0]F(t,s|{\mathbf z}_0),
\label{equation_S}
\end{equation}
where
${\cal L}^*$ is the backward transport operator
\begin{equation}
{\cal L}^* = {\mathbf v}_0\cdot \nabla_{{\mathbf r}_0} + \frac{1}{\tau_s}\int d{\mathbf \omega_0}-\frac{1}{\tau_t}.
\end{equation}

\begin{figure}[t]
\centerline{\epsfxsize=8.0cm\epsfbox{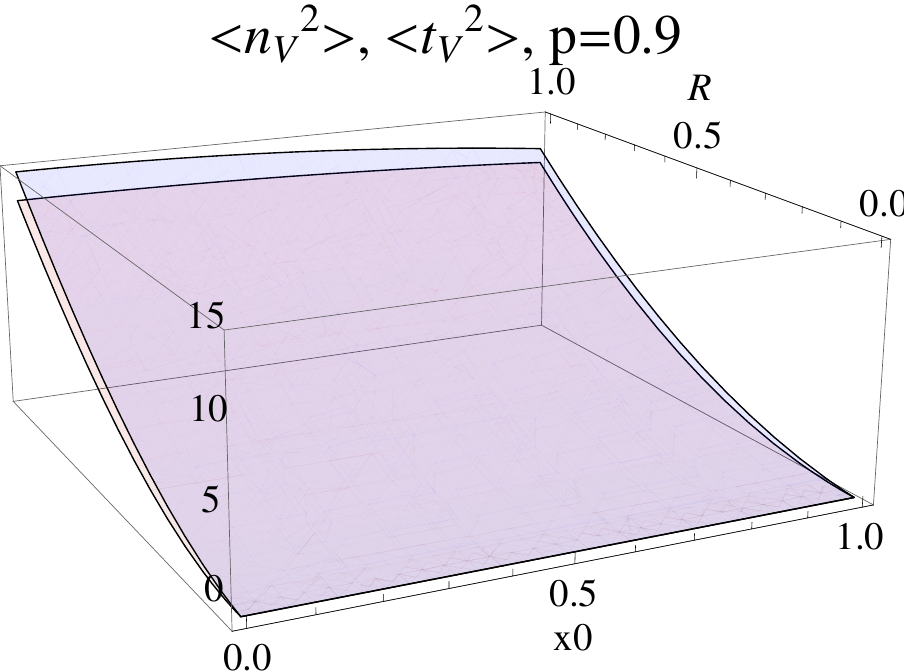}}
\caption{Second moment of residence time (dark surface) and collision number (lights surface) for transparent boundaries and $p=0.9$.}
   \label{fig5}
\end{figure}

Kac has shown that $F(t,s|{\mathbf z}_0)$ can be interpreted as the Laplace transform (the transformed variable being $s$) of ${\cal Q}(t_{V}|{\mathbf z}_0)$. The standard approach would therefore imply first solving Eq.~\eqref{equation_S} for $F(t,s|{\mathbf z}_0)$, and obtaining then ${\cal Q}(t_{V}|{\mathbf z}_0)$ by performing an inverse Laplace transform. Eqs.~\eqref{equation_S} and~\eqref{stochastic_integral} are known as the Feynman-Kac formulae~\cite{majumdar_review}. Once $F(t,s|{\mathbf z}_0)$ is known, the moments of residence time can be obtained from
\begin{equation}
\langle t_V^m \rangle({\mathbf z}_0,t) = (-1)^m \frac{\partial^m}{\partial s^m} F(t,s|{\mathbf z}_0) \vert_{s=0}.
\label{recursion_mom}
\end{equation}
Eqs.~\eqref{equation_S} and~\eqref{recursion_mom} yield the recursion property
\begin{equation}
\frac{\partial}{\partial t}\langle t_V^m \rangle({\mathbf z}_0,t)={\cal L}^* \langle t_V^m \rangle({\mathbf z}_0,t)+m\chi[{\mathbf z}_0]\langle t_V^{m-1} \rangle({\mathbf z}_0,t),
\end{equation}
with the conditions $\langle t_V^m \rangle({\mathbf z}_0,0)=0 $ and $\langle t_V^0 \rangle({\mathbf z}_0,t)=1 $. In most applications, the observation time is assumed to be infinite, i.e., $t\to +\infty$, which leads to the simplified equation
\begin{equation}
{\cal L}^* \langle t_V^m \rangle({\mathbf z}_0)=-m\chi[{\mathbf z}_0]\langle t_V^{m-1} \rangle({\mathbf z}_0),
\label{L_moments}
\end{equation}
where $t_V({\mathbf z}_0)=\lim_{t\to +\infty}t_V({\mathbf z}_0,t)$. Eq.~\eqref{L_moments} is proposed in~\cite{benichou_epl} to generalize a result by~\cite{mazzolo} and derive an elegant recursion formula for the moments of residence (and first-passage) times of exponential fights averaged over initial conditions. In the context of Brownian motion, the relevance of Eq.~\eqref{L_moments} is discussed at length in~\cite{agmon}.

\begin{figure}[t]
\centerline{\epsfxsize=8.0cm\epsfbox{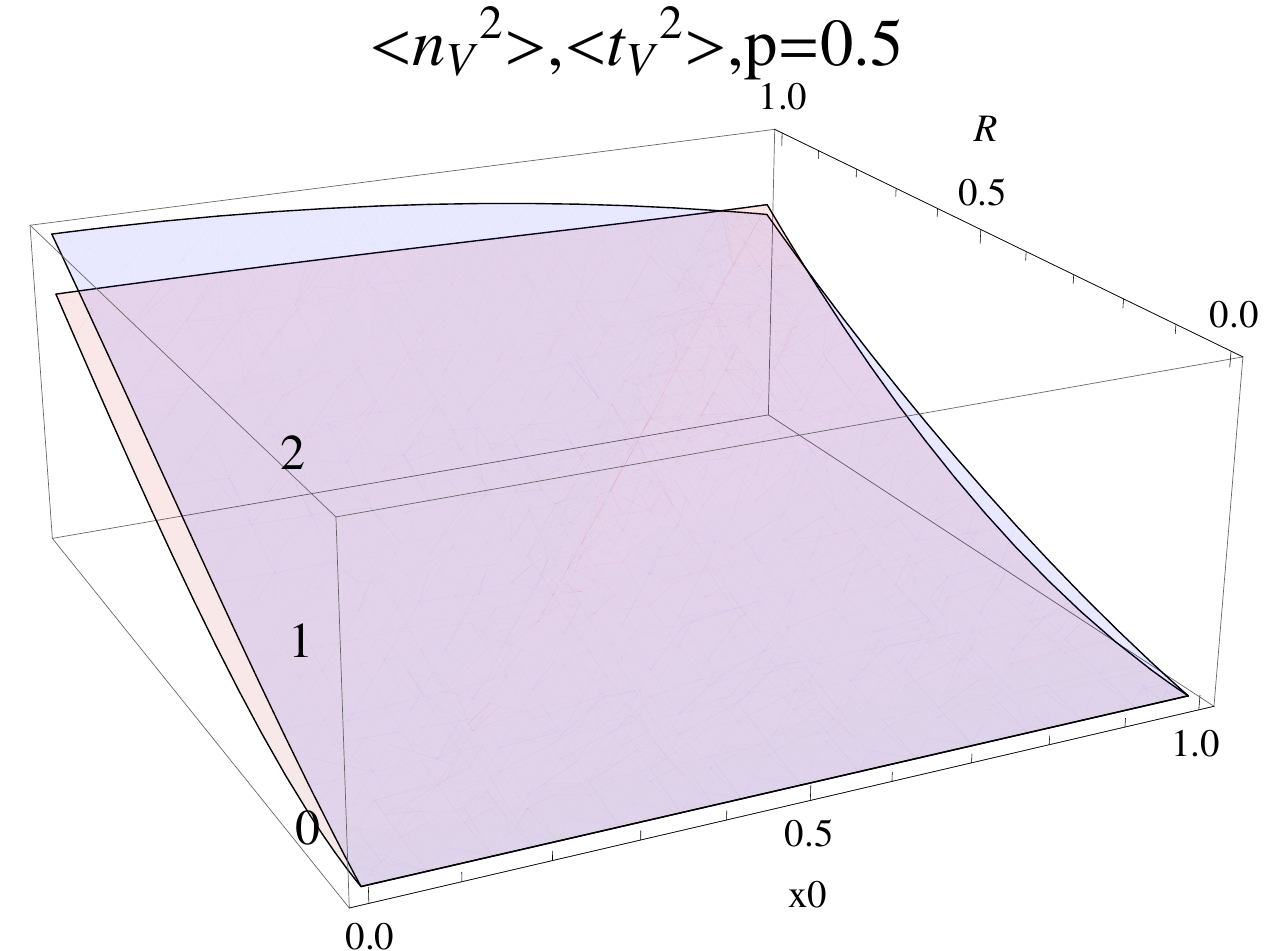}}
\caption{Second moment of residence time (dark surface) and collision number (light surface) for transparent boundaries and $p=0.5$.}
   \label{fig6}
\end{figure}

When one is interested only in the moments of the distribution, and the observation time is infinite, the Feynman-Kac formalism may be rather cumbersome (Eq.~\eqref{L_moments} would still require inversing the backward operator ${\cal L}^*$), and can be altogether avoided by resorting to the so-called Kac moment formula~\cite{kac}. This approach has been successfully applied to the study of the residence time of Brownian particles in~\cite{berezhkovskii}. For a review, see, e.g.,~\cite{pitman}. The $m$-th moment of the residence time is obtained from
\begin{widetext}
\begin{equation}
\langle t_V^m \rangle({\mathbf z}_0) = m! \int d{\mathbf z}_m \int_0^{+\infty} dt_m...\int^{t_2}_0 dt_1 \Psi({\mathbf z}_m,t_m-t_{m-1}|{\mathbf z}_{m-1})*...*\Psi({\mathbf z}_1,t_1|{\mathbf z}_{0}),
\label{kac_t}
\end{equation}
\end{widetext}
where the convolution products read
\begin{widetext}
\begin{equation}
\Psi({\mathbf z}_{i+1},t_{i+1}-t_i|{\mathbf z}_{i})*\Psi({\mathbf z}_i,t_i-t_{i-1}|{\mathbf z}_{i-1}) = \int d{\mathbf z}_i \Psi({\mathbf z}_{i+1},t_{i+1}-t_i|{\mathbf z}_{i})\Psi({\mathbf z}_i,t_i-t_{i-1}|{\mathbf z}_{i-1}).
\end{equation}
\end{widetext}
Finally, by interchanging the order of integration in time, and extending the integration limit to infinity for each convolution product~\cite{berezhkovskii}, we have the formula for the moments of the residence time, namely,
\begin{equation}
\frac{\langle t_V^m \rangle({\mathbf z}_0)}{\tau_t^m} = m! \int d{\mathbf z}_m \Psi({\mathbf z}_m|{\mathbf z}_{m-1})*...*\Psi({\mathbf z}_1|{\mathbf z}_{0}),
\label{eq_moment_t}
\end{equation}
where
\begin{equation}
\Psi({\mathbf z}_{i+1}|{\mathbf z}_{i})= \frac{1}{\tau_t}\int_0^{+\infty}\Psi({\mathbf z}_{i+1},t|{\mathbf z}_{i}) dt
\end{equation}
is the collision density. Eq.~\eqref{eq_moment_t} thus allows expressing the moments of the residence time as a function of the particle equilibrium distribution.

\begin{figure}[t]
\centerline{\epsfxsize=8.0cm\epsfbox{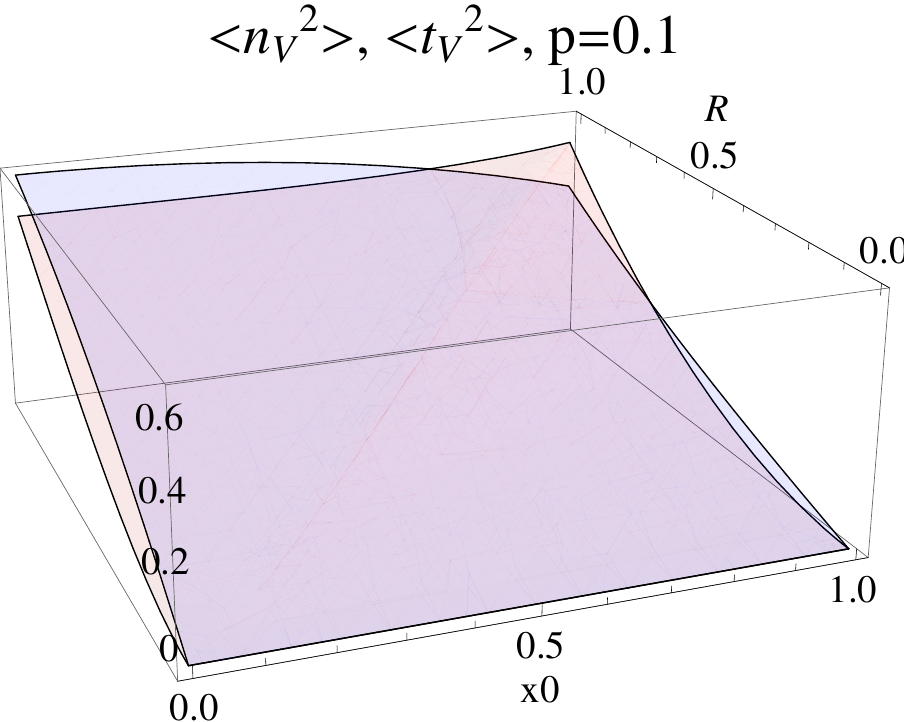}}
\caption{Second moment of residence time (dark surface) and collision number (light surface) for transparent boundaries and $p=0.1$.}
   \label{fig7}
\end{figure}

\subsection{Collision number}
\label{collision_sec}

In a previous work~\cite{zdm_prl}, we have explicitly derived the moments of the collision number $n_V$ for a broad class of renewal processes, when the point-source emits isotropically. For exponential flights, it is sufficient to remark that the process ${\mathbf r}_n$, i.e., the direction-averaged position of the walker, is Markovian at each collision event. The probability of performing $n_V$ collisions in the volume $V$ is related to the propagator by
\begin{eqnarray}
{\cal P}(n_{V}|{\mathbf r}_0)= \int d {\mathbf r} \Psi({\mathbf r},n_{V}|{\mathbf r}_0)-
\int d {\mathbf r} \Psi({\mathbf r},n_{V}+1|{\mathbf r}_0).
\label{pnv}
\end{eqnarray}
We introduce the direction-averaged collision density
\begin{eqnarray}
\Psi({\mathbf r}|{\mathbf r}_0)=\sum_{n=1}^{+\infty} \Psi({\mathbf r},n|{\mathbf r}_0).
\end{eqnarray}
The derivation of the moments $\langle n^m_V \rangle({\mathbf r}_0)$ closely follows that of $\langle t^m_V \rangle({\mathbf z}_0)$~\cite{zdm_prl}. Here we just recall that the moments of $n_V$ are given by
\begin{equation}
\langle n_V^m \rangle({\mathbf r}_0) = \frac{1}{p} \sum_{k=1}^{m} s_{m,k} p^k {\cal C}_k({\mathbf r}_0),
\label{n_moments}
\end{equation}
where the coefficients
\begin{equation}
s_{m,k}=\frac{1}{k!} \sum_{i=0}^{k} (-1)^i \binom {k} {i} \left( k-i\right)^m
\end{equation}
are the Stirling numbers of second kind~\cite{erdelyi}, and
\begin{equation}
{\cal C}_k({\mathbf r}_0) = k! \int_{V}d {\mathbf r}_k ... \int_{V} d {\mathbf r}_1 \Psi({\mathbf r}_k|{\mathbf r}_{k-1})...\Psi({\mathbf r}_1|{\mathbf r}_0)
\label{kac_integrals}
\end{equation}
are defined as $k$-fold convolutions of the collision density $\Psi({\mathbf r}|{\mathbf r}_0)$ with itself~\cite{zdm_prl}.

Now, the moments $\langle n_V^m \rangle({\mathbf r}_0,{\mathbf \omega}_0)$ for a directed source $\delta\left({\mathbf r}-{\mathbf r}_0 \right) \delta\left( {\mathbf \omega}-{\mathbf \omega}_0\right)$ can be evaluated as follows. First, we compute the density $\pi({\mathbf r}'|{\mathbf r}_0,{\mathbf \omega}_0)$ of the walkers entering their first collision at ${\mathbf r}'$. Each first-collision point will re-emit isotropically after the collision, i.e., the distribution of the outgoing ${\mathbf \omega}'$ is uniform. Then, the moments $\langle n_V^m \rangle({\mathbf r}_0,{\mathbf \omega}_0)$ are obtained by convoluting Eq.~\eqref{n_moments} for an isotropic source at ${\mathbf r}'$ with the first-collision source $p\pi({\mathbf r}'|{\mathbf r}_0,{\mathbf \omega}_0)$. In terms of collision number probabilities, we have
\begin{widetext}
\begin{equation}
{\cal P}(n_{V}|{\mathbf r}_0,{\mathbf \omega}_0)=p\int d{\mathbf r}' \chi[{\mathbf r}'] {\cal P}(n_{V}-1|{\mathbf r}')\pi({\mathbf r}'|{\mathbf r}_0,{\mathbf \omega}_0)+p\int d{\mathbf r}' \tilde{\chi}[{\mathbf r}'] {\cal P}(n_{V}|{\mathbf r}')\pi({\mathbf r}'|{\mathbf r}_0,{\mathbf \omega}_0),
\end{equation}
\end{widetext}
where $\tilde{\chi}[{\mathbf r}']$ vanishes for ${\mathbf r}' \in V$ and is equal to one elsewhere. This leads to
\begin{widetext}
\begin{equation}
\langle n_V^m \rangle({\mathbf r}_0,{\mathbf \omega}_0)=p\sum_{k=0}^{m-1}\binom{m}{k}\int d{\mathbf r}' \chi[{\mathbf r}'] \langle n_V^k \rangle({\mathbf r}')\pi({\mathbf r}'|{\mathbf r}_0,{\mathbf \omega}_0)+p\int d{\mathbf r}'\langle n_V^m \rangle({\mathbf r}')\pi({\mathbf r}'|{\mathbf r}_0,{\mathbf \omega}_0).
\label{n_moments_def}
\end{equation}
\end{widetext}
For the average collision number, from Eq.~\eqref{n_moments_def} we have in particular
\begin{equation}
\langle n_V^1 \rangle({\mathbf r}_0,{\mathbf \omega}_0)=\int d{\mathbf r}' \left[\chi[{\mathbf r}']+ p\int d{\mathbf r} \Psi({\mathbf r}|{\mathbf r}')\right]\pi({\mathbf r}'|{\mathbf r}_0,{\mathbf \omega}_0).
\end{equation}
By remarking that
\begin{equation}
\chi({\mathbf r})\pi({\mathbf r}|{\mathbf r}_0,{\mathbf \omega}_0)+ p\int d{\mathbf r}' \Psi({\mathbf r}|{\mathbf r}')\pi({\mathbf r}'|{\mathbf r}_0,{\mathbf \omega}_0)=\Psi({\mathbf r}|{\mathbf r}_0,{\mathbf \omega}_0),
\end{equation}
we therefore get
\begin{equation}
\langle n_V^1 \rangle({\mathbf r}_0,{\mathbf \omega}_0)=\int d{\mathbf r} \Psi({\mathbf r}|{\mathbf r}_0,{\mathbf \omega}_0).
\end{equation}
Observe that for the mean residence time we have
\begin{equation}
\frac{\langle t_V^1 \rangle({\mathbf r}_0,{\mathbf \omega}_0)}{\tau_t} = \int d{\mathbf r}\int d{\mathbf \omega} \Psi({\mathbf r},{\mathbf \omega}|{\mathbf r}_0,{\mathbf \omega}_0),
\end{equation}
Since $\Psi({\mathbf r}|{\mathbf r}_0,{\mathbf \omega}_0)=\int d{\mathbf \omega} \Psi({\mathbf r},{\mathbf \omega}|{\mathbf r}_0,{\mathbf \omega}_0)$, it finally follows the (rescaled) equality between the average residence time and the average collision number, namely, $\langle n_V^1 \rangle({\mathbf r}_0,{\mathbf \omega}_0)=\langle t_V^1 \rangle({\mathbf r}_0,{\mathbf \omega}_0)/\tau_t$.

\begin{figure}[t]
\centerline{\epsfxsize=8.0cm\epsfbox{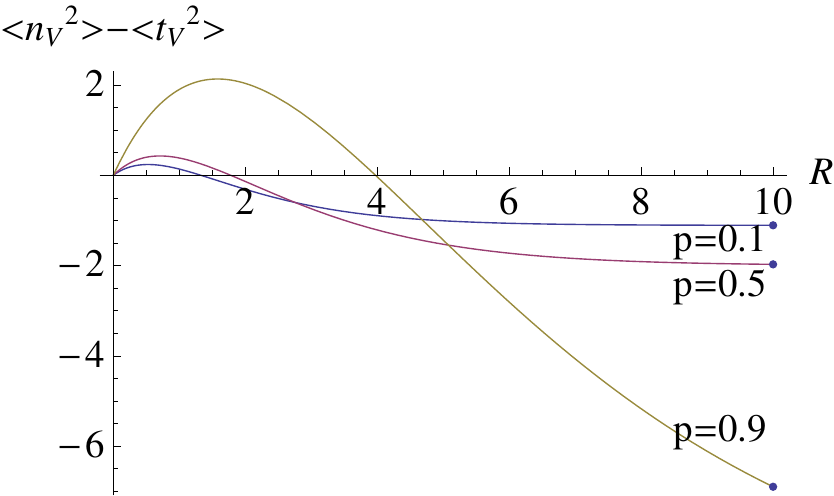}}
\caption{The quantity $\langle n_V^2 \rangle({\mathbf r}_0)-\langle t_V^2 \rangle({\mathbf x}_0)$ when ${\mathbf x}_0=0$, as a function of $R$ and the scattering probability $p$.}
   \label{fig8}
\end{figure}

\section{The rod problem}
\label{rod_sec}

The approach presented in the previous Section allows explicitly evaluating the moments $\langle n_V^m \rangle({\mathbf r}_0,{\mathbf \omega}_0)$ and $\langle t_V^m \rangle({\mathbf r}_0,{\mathbf \omega}_0)$. When the equilibrium distribution is known, this amounts to solving the convolution integrals in Eqs.~\eqref{n_moments_def} and~\eqref{eq_moment_t}, respectively. However, analytical expressions for $\Psi({\mathbf r},{\mathbf \omega}|{\mathbf r}_0,{\mathbf \omega}_0)$ or $\Psi({\mathbf r}|{\mathbf r}_0)$ (subject to the appropriate boundary conditions) are known only in a few cases~\cite{zoia_dumonteil_mazzolo, zdm_prl}, so that one must generally resort to numerical integration. A well-known and long-studied example where calculations can be carried out analytically is the so-called {\em rod model}, where particles can move along a straight line~\cite{wing, weiss, hughes}. This corresponds to exponential flights in $1d$, with only forward and backward direction allowed. Though the rod model is somewhat inadequate to address realistic radiation transport phenomena, we shall discuss it here for two main reasons. First, it allows illustrating the application of the above formulas for $\langle n_V^m \rangle({\mathbf r}_0,{\mathbf \omega}_0)$ and $\langle t_V^m \rangle({\mathbf r}_0,{\mathbf \omega}_0)$, and provides some hints on their use as Monte Carlo estimators. Second, the rod model, despite being admittedly oversimplified, is nonetheless widely used in biology (often called velocity jump process), gas dynamics (Lorentz gas), finance and neutronics, as it captures the essential features of the corresponding physical system~\cite{velocity_jump, bacteria, weiss_review, wing, lorentz}.

We define $\omega_f$ and $\omega_b$ the forward and backward directions, respectively. Similarly, we denote by $S_f$ and $S_b$ the forward and backward components of the source, located at $x_0$. Furthermore, we denote by $x$ the abscissa of the rod, positive when oriented as $\omega_f$. We set the mean free path $\lambda_t=1$, and we take $v=1$. Scattering is isotropic. The volume $V$ is assumed to be the interval $[-R,R]$. With this choice of parameters and notations, Eq.~\eqref{chapman_kolmogorov} reduces to the following set of stationary first-order differential equations
\begin{widetext}
\begin{eqnarray}
\left( \frac{\partial}{\partial x} +1\right) \Psi(x,\omega_f|x_0,\omega_f)=\frac{p}{2}\left[ \Psi(x,\omega_f|x_0,\omega_f)+\Psi(x,\omega_b|x_0,\omega_f)\right]  + S_f\nonumber \\
\left( -\frac{\partial}{\partial x} +1\right) \Psi(x,\omega_b|x_0,\omega_f)=\frac{p}{2}\left[\Psi(x,\omega_b|x_0,\omega_f)+\Psi(x,\omega_f|x_0,\omega_f) \right]
\label{eq_rod_f}
\end{eqnarray}
\end{widetext}
when the source is $S_f$, and
\begin{widetext}
\begin{eqnarray}
\left( \frac{\partial}{\partial x} +1\right) \Psi(x,\omega_f|x_0,\omega_b)=\frac{p}{2}\left[\Psi(x,\omega_f|x_0,\omega_b)+\Psi(x,\omega_b|x_0,\omega_b) \right] \nonumber \\
\left( -\frac{\partial}{\partial x} +1\right) \Psi(x,\omega_b|x_0,\omega_b)=\frac{p}{2}\left[\Psi(x,\omega_b|x_0,\omega_b)+\Psi(x,\omega_f|x_0,\omega_b) \right] + S_b
\label{eq_rod_b}
\end{eqnarray}
\end{widetext}
when the source is $S_b$.

Two relevant examples will be considered here: $i)$ leakage boundary conditions (a first-passage problem) without absorption, and $ii)$ transparent boundaries with absorption. In the former case, leakages at $x=\pm R$ impose $\Psi(-R,\omega_f|x_0,\omega_f)=0$, $\Psi(R,\omega_b|x_0,\omega_f)=0$, $\Psi(R,\omega_b|x_0,\omega_b)=0$, and $\Psi(-R,\omega_f|x_0,\omega_b)=0$, which corresponds to an homogeneous medium surrounded by vacuum. The source is therefore $x_0 \in V$. In the latter, boundary conditions are imposed at infinity, which corresponds to an infinite homogeneous medium, the boundaries of $V$ being transparent and not affecting particle trajectories. One-dimensional exponential flights are recurrent walks (i.e., they almost surely re-visit their initial position)~\cite{zoia_dumonteil_mazzolo, zdm_prl}, so that it is necessary to impose leakages and/or set $p<1$ in order to prevent $\langle n_V^m \rangle({\mathbf x}_0,{\mathbf \omega}_0)$ and $\langle t_V^m \rangle({\mathbf x}_0,{\mathbf \omega}_0)$ from diverging.

For the case of purely scattering media, i.e., $p=1$, and leakage boundaries, the rod problem equations~\eqref{eq_rod_f} and~\eqref{eq_rod_b} are straightforwardly solved by direct integration, and give rise to first-order discontinuous polynomials, the discontinuity being located at $x_0$, i.e., at the source. Once the four solutions $\Psi(x,\omega_f|x_0,\omega_f)$, $\Psi(x,\omega_b|x_0,\omega_f)$, $\Psi(x,\omega_b|x_0,\omega_b)$, and $\Psi(x,\omega_f|x_0,\omega_b)$ have been obtained, the moments $\langle n_V^m \rangle({\mathbf x}_0)$ and $\langle t_V^m \rangle({\mathbf x}_0)$ are computed by performing the convolution integrals in Eqs.~\eqref{n_moments_def} and~\eqref{eq_moment_t}, respectively. Remark that the isotropic source corresponds to assuming $S_f=S_b$ and integrating with respect to the initial direction. For the mean first-passage time, we have
\begin{equation}
\langle t_V^1 \rangle({\mathbf x}_0)=\frac{R^2+ 2R - x_0^2}{2},
\end{equation}
with $\langle n_V^1 \rangle({\mathbf x}_0)=\langle t_V^1 \rangle({\mathbf x}_0)$ ($v=1$, so that $\tau_t=1$). For the second moment, we have
\begin{widetext}
\begin{equation}
\langle n_V^2 \rangle({\mathbf x}_0) = \frac{12 R + 30 R^2 + 20 R^3 + 5 R^4 - 6 x_0^2 - 12 R x_0^2 -6 R^2 x_0^2 + x_0^4}{12}
\end{equation}
\end{widetext}
and
\begin{widetext}
\begin{equation}
\langle t_V^2 \rangle({\mathbf x}_0) = \frac{12 R^2 + 20 R^3 + 5 R^4 + 12 x_0^2 - 12 R x_0^2 - 6 R^2 x_0^2 + x_0^4}{12}.
\end{equation}
\end{widetext}
The terms in these formulas look inhomogeneous (this is due to setting $\lambda_t=1$), but expressions are indeed dimensionless. The surfaces are discontinuous, since $|x_0| \le R$. Observe that when $R$ is large we have $\langle t_V^2 \rangle({\mathbf x}_0)\simeq\langle n_V^2 \rangle({\mathbf x}_0)$. When $R\to+\infty$, the moments diverge, as expected from $1d$ exponential flights being recurrent walks. Observe that the first and second moment of the first-passage time satisfy the recursion property derived in~\cite{benichou_epl}, namely
\begin{equation}
\left\lbrace \langle t_V^{m-1} \rangle({\mathbf x}_0) \right\rbrace_V = \frac{\left\lbrace \langle t_V^{m} \rangle({\mathbf x}_0) \right\rbrace_\Sigma}{m\left\lbrace \langle t_V^{1} \rangle({\mathbf x}_0) \right\rbrace_\Sigma}
\label{benichou_average}
\end{equation}
for $m\ge 1$, where $\left\lbrace \cdot \right\rbrace_\Sigma$ and $\left\lbrace \cdot \right\rbrace_V$ denote averaging ${\mathbf x}_0$ over the surface $\Sigma$ (of $V$) or the volume $V$, respectively. Here $d=1$ and $m=2$, and it is easy to verify that
\begin{eqnarray}
\left\lbrace \langle t_V^1 \rangle({\mathbf x}_0) \right\rbrace_\Sigma = R \nonumber\\
\left\lbrace \langle t_V^1 \rangle({\mathbf x}_0) \right\rbrace_V = R+\frac{1}{3}R^2 \nonumber\\
\left\lbrace \langle t_V^2 \rangle({\mathbf x}_0) \right\rbrace_\Sigma = 2\left( R^2+\frac{1}{3}R^3\right) ,
\end{eqnarray}
hence $\left\lbrace \langle t_V^1 \rangle({\mathbf x}_0) \right\rbrace_V = \left\lbrace \langle t_V^2 \rangle({\mathbf x}_0) \right\rbrace_\Sigma / 2\left\lbrace \langle t_V^1 \rangle({\mathbf x}_0) \right\rbrace_\Sigma$. Remark that we have an overall factor $1/2$ with respect to the surface averages in~\cite{benichou_epl}, because trajectories are here allowed starting from $\Sigma$ in the outward direction, whereas in~\cite{benichou_epl} they are not. In Fig.~\ref{fig2} we display the mean collision number $\langle n_V^1 \rangle({\mathbf x}_0)$ and the mean first-passage time $\langle t_V^1 \rangle({\mathbf x}_0)$ for leakage boundary conditions. The two surfaces, as a function of ${\mathbf x}_0$ and $R$, coincide, as expected from the considerations exposed above. This goes along with the collision and track length Monte Carlo estimators being unbiased with respect to each other. The second moments $\langle n_V^2 \rangle({\mathbf x}_0)$ and $\langle t_V^2 \rangle({\mathbf x}_0)$ are displayed in Fig.~\ref{fig3}. It is immediately apparent that $\langle n_V^2 \rangle({\mathbf x}_0) \ge \langle t_V^2 \rangle({\mathbf x}_0)$ (actually, equality is attained only for $R \gg 1$): this means that for this example the use of a track length estimator is to be preferred, as it would lead to a smaller variance. All analytical results have been validated by comparison with Monte Carlo simulations.

For $p<1$ and transparent boundaries, the solutions of the rod problem~\eqref{eq_rod_f} and~\eqref{eq_rod_b} are given by combinations of exponential functions, rather than linear polynomials. In this case, the expressions for the moments are rather cumbersome and will not be reported here. Instead, we plot the moments as a function of the initial condition ${\mathbf x}_0$, the domain size $R$ and the scattering rate $p$. In Fig.~\ref{fig4} we display the mean collision number $\langle n_V^1 \rangle({\mathbf x}_0)$ and the mean residence time $\langle t_V^1 \rangle({\mathbf x}_0)$ for transparent boundaries and $p=0.5$. The two surfaces, as a function of ${\mathbf x}_0$ and $R$, coincide, and this relation holds for any value of $p$. The second moments $\langle n_V^2 \rangle({\mathbf x}_0)$ and $\langle t_V^2 \rangle({\mathbf x}_0)$ are displayed in Figs.~\ref{fig5} ($p=0.9$),~\ref{fig6} ($p=0.5$), and~\ref{fig7} ($p=0.1$). In this case, it is not possible to establish a simple inequality between the two surfaces, independent of $p$. As the scattering rate varies, the surfaces change and there exist a value of $p$ for which $\langle n_V^2 \rangle({\mathbf x}_0)$ is smaller than $\langle t_V^2 \rangle({\mathbf x}_0)$. This means that in presence of absorption the collision estimator may lead to a smaller variance. In Fig.~\ref{fig8} we display the difference $\langle n_V^2 \rangle({\mathbf x}_0)-\langle t_V^2 \rangle({\mathbf x}_0)$ when ${\mathbf x}_0=0$, for various values of $p$: when $R$ is large, $\langle t_V^2 \rangle({\mathbf x}_0)$ becomes larger than $\langle n_V^2 \rangle({\mathbf x}_0)$, and this behavior is enhanced for small values of $p$, i.e., large absorption rates. When $R$ is large, the dependence on the angular variable ${\mathbf \omega}$ gets progressively weaker, so that the integrals~\eqref{kac_integrals} and~\eqref{eq_moment_t} coincide. Under this assumptions, calculations show that $\langle n_V^1 \rangle({\mathbf x}_0) \simeq 1/(1-p)$, and $\langle n_V^2 \rangle({\mathbf x}_0) \simeq (1+p)/(1-p)^2$, independent of the initial condition. Then, from the equality of the Kac convolution integrals, the difference $\langle n_V^2 \rangle({\mathbf x}_0)-\langle t_V^2 \rangle({\mathbf x}_0)$ for large $R$ converges to the limit $1/(p-1)$, which implies a smaller variance for the collision estimator. We have also verified that $\langle t_V^1 \rangle({\mathbf x}_0)$ and $\langle t_V^2 \rangle({\mathbf x}_0)$ satisfy the surface- and volume-averaged recursion property in~\cite{benichou_epl}, which generalizes Eq.~\eqref{benichou_average} to residence times. Again, all analytical results have been validated by comparison with Monte Carlo simulations.

\section{Conclusions}
\label{conclusions}

Motivated by their relevance for stochastic transport phenomena as well as for Monte Carlo methods, in this paper we have examined the moments of collision number $n_V$ and residence time $t_V$ of exponential flights in a volume $V$. We have presented a general approach that - based on the Kac moments formula - allows explicitly evaluating such quantities in terms of repeated convolutions of the particle equilibrium distribution. To exemplify the proposed formalism, we have in particular analyzed a $1d$ system, the so-called rod problem, where closed form expressions can be found. We have therefore explicitly computed the moments of $n_V$ and $t_V$ for various boundary conditions, focusing in particular on the first and second moment. Finally, the relevance of these findings in the context of Monte Carlo collision and track length estimators has been discussed. Results show that the averages of $n_V$ and $t_V$ coincide, whereas the second moments (hence the variances) depend on boundary and initial conditions, and on the scattering probability. Residence time has in general a smaller variance, but the opposite is true when absorption dominates over scattering.

By virtue of the increasing power of Monte Carlo methods in solving realistic three-dimensional transport problems, one might argue that such a simple system as the rod problem has a limited interest. On the contrary, we are persuaded that this analysis is useful, in that it allows focusing on the essential features of the physical system at hand. Indeed, on one hand it sheds light at the deep connections between sojourn times and collision number for exponential flights, and on the other hand it gives some hints on the behavior of the intrinsic variance of Monte Carlo collision and track length estimators. Extending the proposed approach to higher-dimensional and more complex systems is highly desirable, and investigations to this aim are ongoing.

\acknowledgments

The authors wish to thank Dr.~F.~Malvagi for useful discussions and T.~Lefebvre for help in preparing figures.

\end{document}